\documentclass[10pt, conference, compsocconf]{IEEEtran}
\usepackage{cite}
\usepackage{indentfirst}
\usepackage{amsfonts}
\usepackage{latexsym}
\usepackage{bm}
\usepackage{amssymb}

\usepackage{algorithm}
\usepackage{multirow} 
\usepackage{amsmath}
\usepackage{xcolor}
\usepackage{algpseudocode}
\usepackage{subfigure}
\usepackage{hyperref}

\ifCLASSINFOpdf
\usepackage[pdftex]{graphicx}
\else
\fi
\begin{document}

\renewcommand{\algorithmicrequire}{\textbf{Initialization:}}
\renewcommand{\algorithmicensure}{\textbf{Output:}}
\newcommand{\tabincell}[2]{\begin{tabular}{@{}#1@{}}#2\end{tabular}}
%
\title{Service Capacity Enhanced Task Offloading and Resource Allocation
in Multi-Server Edge Computing Environment}
\author{
\IEEEauthorblockN{
  Wei Du\IEEEauthorrefmark{1}\IEEEauthorrefmark{2},
  Tao Lei\IEEEauthorrefmark{1},
  Qiang He\IEEEauthorrefmark{3},
  Wei Liu\IEEEauthorrefmark{1}\IEEEauthorrefmark{2},
  Qiwang Lei\IEEEauthorrefmark{1},
  Hailiang Zhao\IEEEauthorrefmark{1},
  Wei Wang\IEEEauthorrefmark{4}
  }
\IEEEauthorblockA{
  \IEEEauthorrefmark{1}
  Wuhan University of Technology, Wuhan, China\\
  \{whutduwei,leitao1995, wliu, leiqiwang\} @whut.edu.cn,
  hliangzhao97@gmail.com
}
\IEEEauthorblockA{
  \IEEEauthorrefmark{2}
  Hubei Key Laboratory of Transportation Internet of Things, Wuhan, China
}
\IEEEauthorblockA{
  \IEEEauthorrefmark{3}
  Swinburne University of Technology, Melbourne, Australia\\
  qhe@swin.edu.au
}
\IEEEauthorblockA{
  \IEEEauthorrefmark{4}
  East China Normal University, ShangHai, China\\
  wwang@dase.ecnu.edu.cn
}
}
\maketitle
\begin{abstract}
  An edge computing environment features multiple edge servers and multiple service clients. In this environment, mobile service providers can offload client-side computation tasks from service clients' devices onto edge servers to reduce service latency and power consumption experienced by the clients. A critical issue that has yet to be properly addressed is how to allocate edge computing resources to achieve two optimization objectives: 1) minimize the service cost measured by the service latency and the power consumption experienced by service clients; and 2) maximize the service capacity measured by the number of service clients that can offload their computation tasks in the long term. This paper formulates this long-term problem as a stochastic optimization problem and solves it with an online algorithm based on Lyapunov optimization. This NP-hard problem is decomposed into three sub-problems, which are then solved with a suite of techniques. The experimental results show that our approach significantly outperforms two baseline approaches.
\end{abstract}
%
\IEEEpeerreviewmaketitle
\section{Introduction}
\indent Edge computing has emerged as a new paradigm for powering applications by offering computing, storage and networking resources at the edge of the cloud \cite{mao2017survey, ZhouYuNearEnd2018}. As illustrated on the 5G standardization roadmap, edge servers will be distributed at ultra-dense small-cell base stations (CBS) \cite{ge20165g, QiYanli478}. In such an environment, the coverages of adjacent edge servers partially overlap to avoid blank areas not covered by any edge servers \cite{lai2018optimal}. Service providers can offload client-side computation tasks from service clients' devices onto edge servers to reduce the service cost measured by the service latency and power consumption experienced by service clients \cite{zhang2018qcss, wu2018service}. This is referred to as \textit{task offloading}.

The edge infrastructure provider looks at task offloading from
two general perspectives. A large number of edge servers sharing
service clients’ offloaded computation tasks can provide service clients with low service cost from service clients' perspective. That is the benefit of computation loading. However, the deployment of excessive edge servers result in overly high operational cost from edge infrastructure provider's perspective. After all, the edge infrastructure provider is interested in the overall revenue, i.e., the benefit minus the cost. The economy of scale maximizes the edge infrastructure provider’s revenue by maximizing the service capacity,i.e., the ability to serve the maximum number of service clients. To achieve a cost-effective solution, the service cost needs to be traded off to allow more service clients to be served by edge servers. Thus, how to trade off the service cost and service capacity is a critical problem in edge computing.

A lot of researchers have focused on solving the task offloading problem in a single-server edge computing environment. However, in a particular area, there are usually multiple edge servers available for offloading service clients' computation tasks. The edge computing environment is in fact a multi-server environment. In such an environment, it is critical and challenging to optimize the utilization of edge computing resources, including computation and transmission resources, with the aim to maximize the service capacity while minimizing the service cost. Firstly, the characteristics of latency-tolerant applications, i.e., the coupling among randomly-arrived tasks, must be captured \cite{mao2017stochastic}. Stochastic computation partitioning strategies must be formulated to split the resources to be shared among service clients. Secondly, each service client needs to decide not only how to partition its computation tasks between its device and the edge server but also which edge server to offload its computation tasks to.

In this paper, we propose a holistic solution to computation
offloading and resource allocation in multi-server edge computing
environment with the aim to serve as many service clients as possible with
minimum service cost. The main contributions of this paper are as
follows:
\begin{itemize}
\item Task offloading in such an environment is modelled as a
stochastic optimization problem with multiple optimization objectives.
It aims to maximize the number of service clients served with minimum service cost in the long term.
\item Based on Lyapunov optimization, the above long-term stochastic optimization
problem is converted to a deterministic optimization problem within each
time slot, which is then further decomposed into three sub-problems.
\item To solve the optimization problem, an online joint task offloading
and resource allocation algorithm (OJTORA) powered by a suite of techniques is proposed to solve each sub-problem with low complexity.
\item Extensive experiments are conducted to evaluate OJTORA. The results show that OJTORA outperforms the baseline approaches significantly.
\end{itemize}

The remainder of this paper is organized as follows. Section
\uppercase\expandafter{\romannumeral2} reviews related work. Section
\uppercase\expandafter{\romannumeral3} presents the edge computing model. Section
\uppercase\expandafter{\romannumeral4} formulates the research problem.
Section \uppercase\expandafter{\romannumeral5} introduces the proposed
approach. Section \uppercase\expandafter{\romannumeral6} experimentally
evaluates the proposed approach. Section \uppercase\expandafter{\romannumeral7}
concludes this paper.

\section{Related Works}

In recent years, joint task offloading and resource allocation in edge computing
has attracted many researchers’ attention. Most existing work studied the problem in a single-server edge computing environment \cite{mao2017stochastic, sardellitti2015joint, lyu2017multiuser, YuBowen:537}.
Some researchers have considered multi-server scenarios \cite{tran2017joint, liu2017latency, masoudi2017green, yang2017network, he2018s, li2017efficient, wang2017online}.
The authors of \cite{tran2017joint} optimize task offloading, uplink
transmission power of service clients, and computing resource allocation on
edge servers to minimize task completion time and power consumption. In
\cite{liu2017latency}, the tradeoff between latency and power consumption
was studied. The problem was formulated as computation and transmits power minimization subject to latency and reliability constraints. The authors of \cite{masoudi2017green} studied how to minimize mobile power consumption through data offloading.
Centralized and distributed algorithms for power allocation and
transmission channel assignment were proposed. In \cite{yang2017network},
device-edge-cloud edge was investigated. A network-aware multi-user and
multi-edge computation partitioning problem was formulated. Computation
and radio transmission resources were allocated such that service clients’
average throughput was maximized. In \cite{he2018s}, the shareable storage was considered, and a constant-factor approximation algorithm was proposed to decide the server placement and resource allocation. In \cite{li2017efficient}, the edge task offloading control was investigated in cloud radio access network (C-RAN) environments,
and a multi-stage heuristic was proposed to minimize the refusal rate for
user’s task offloading requests. In \cite{wang2017online}, an online
algorithm was proposed to allocate edge server's resource over time.

In the studies of task offloading, service latency and power consumption
have been commonly acknowledged as two very important optimization objectives. However, few researchers have considered service capacity, i.e., the total number of service clients served, which is a key issue from the edge infrastructure provider’s perspective in a multi-server edge computing environment\cite{lai2018optimal}. In \cite{alfayly2015qoe}. A quality-of-experience (QoE) driven LTE downlink scheduling scheme for VoIP applications was proposed to improve the service capacity with acceptable QoE for all service clients. However, they only consider single-server environments. In addition, the proposed LTE downlink scheduling scheme is specifically designed for VoIP applications and thus is not applicable to most other applications in the edge computing environment.

Our work differentiates from existing work in the following ways. First, we consider a multi-server edge computing environment. Second, our approach is applicable to most latency-tolerant applications in the edge computing environment. Third, we consider both resource allocation and task offloading. Finally, we attempt to achieve the optimization objectives in the long term.

\section{System Model}

\begin{figure}[!ht]
  \centering
  \includegraphics[width=2.5in]{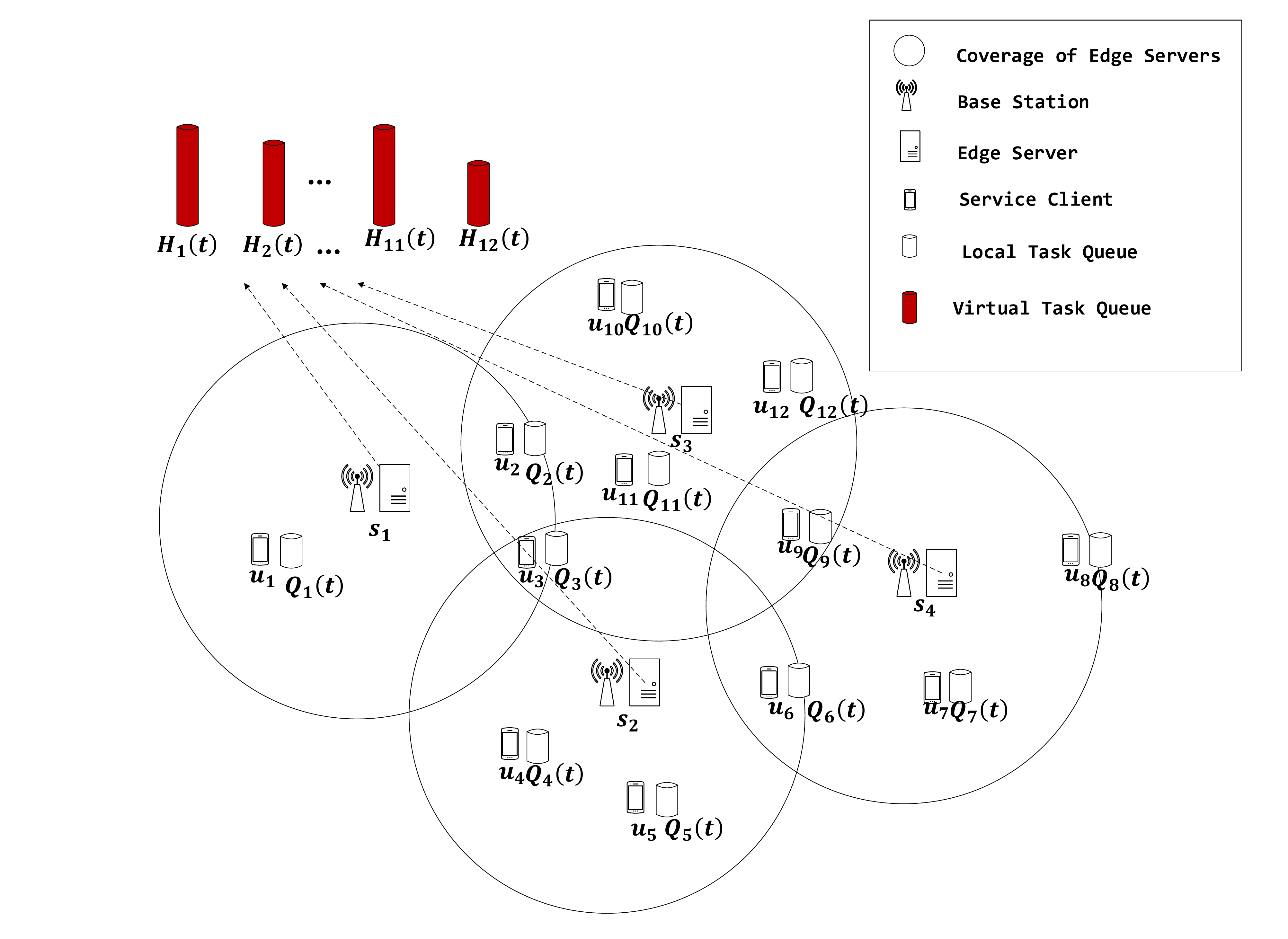}
  \caption{A multi-server edge environment}
  \label{edge}
\end{figure}

\indent An example multi-server edge computing scenario is shown in Fig. \ref{edge}.
A service client can offload some or all of its computation tasks to one of its nearby edge servers.
Let us denote the set of service clients with $\emph{U}$, the set of edge servers with $\emph{S}$, the connection between the \emph{i}th service client and the \emph{j}th edge server with $c_{i,j} = \{0,1\}$, where $c_{i,j} = 1$ indicates that the \emph{i}th service client can access the \emph{j}th edge server or $c_{i,j} = 0$ otherwise.
Let us define
$G_{i} = \{j | c_{i,j} = 1, j \in S\}, i \in U$,
and $Z_{j} = \{i | c_{i,j} = 1, i \in U\}, j \in S$. Task offloading in the edge computing environment is an ongoing process. Let us denote different time slots
with $T = \{1,2,3,...\}$ and the time slot length is $\tau$. The available
bandwidth of each edge server is $\omega$ Hz and the noise power spectral
density  is $N_{0}$.

\subsection{Computation Task and Task Queue Model}

\begin{figure}[!ht]
  \centering
  \includegraphics[width=2.5in]{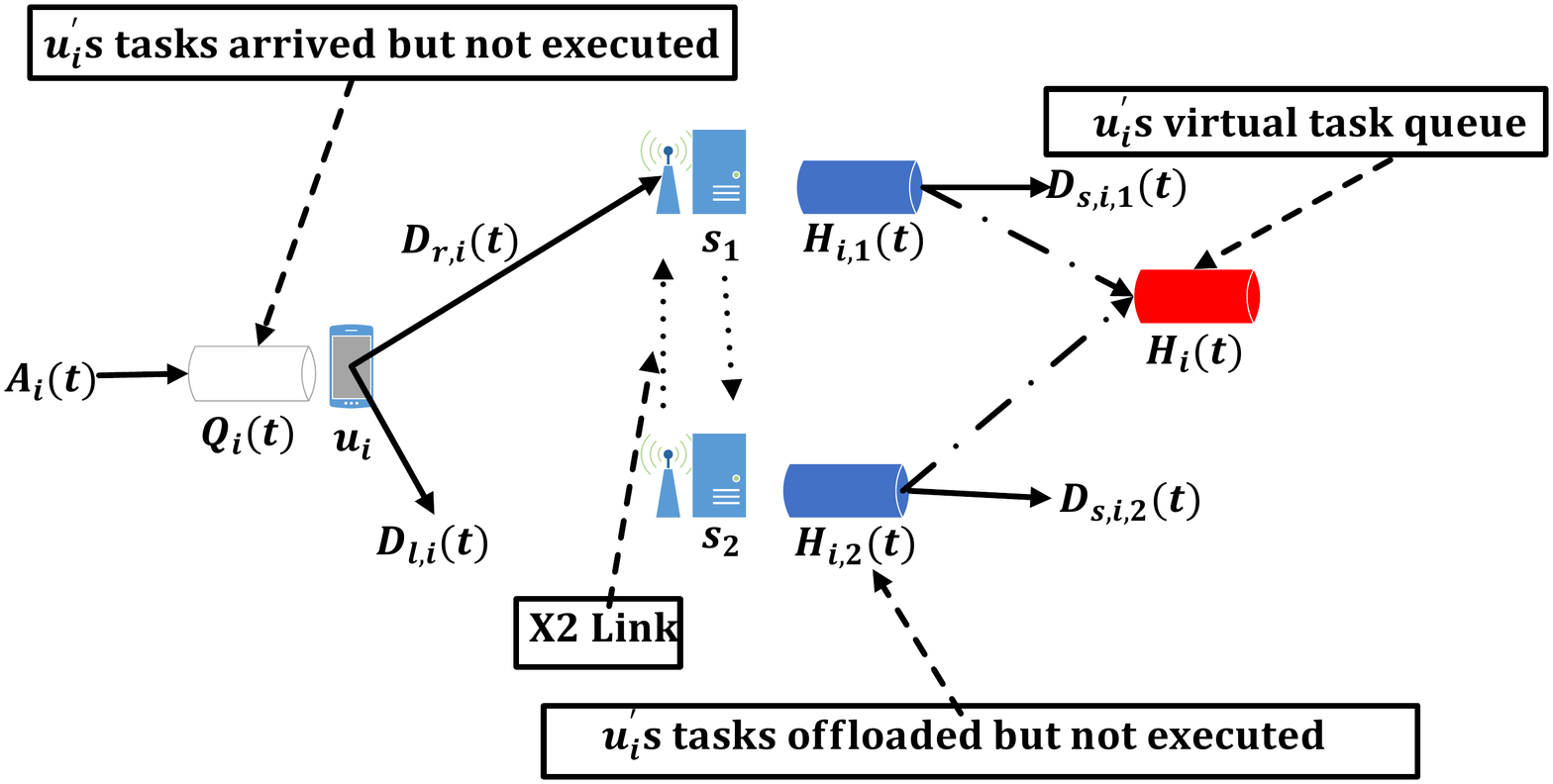}
  \caption{The task queues of edge system}
  \label{VirtualQueue}
\end{figure}

The computation tasks running on service clients' devices are bit-wise independent
\cite{mao2017survey}. Let us denote the amount of tasks of the \emph{i}th service client in the \emph{t}th time slot as $A_{i}(t)$,
which are independent and
identically distributed in different time slot within $[0,A_{i,max}], A_{i,max} \in \mathbb{R}^{+}$,
with the expectation $\mathbb{E}[A_{i}(t)] = \lambda_{i}, \lambda_{i} \in [0,A_{i,max}], i \in U$.
As shown in Fig. \ref{VirtualQueue}, at the beginning of the \emph{t}th time slot, the local queue length of the \emph{i}th service client is $Q_{i}(t)$.
At $t$th time slot, the amount of tasks which has arrived but not been executed locally or offloaded will be put in $Q_{i}(t)$.
Within the \emph{t}th time slot, the \emph{i}th service client will locally execute $D_{l,i}(t)$ tasks
and will offload $D_{r,i}(t)$ tasks to an edge server, i.e.,
$D_{\sum,i}(t)=D_{l,i}(t)+D_{r,i}(t)$.
\begin{equation}
  \centering
  Q_{i}(t+1) = max\{Q_{i}(t) - D_{\sum,i}(t), 0\} + A_{i}(t), i \in U.
  \label{equ1}
\end{equation}
\indent As shown in Fig. \ref{VirtualQueue}, a service client chooses only
one edge server to offload its tasks in each time slot. Thus, we maintain a
virtual task queue $H_{i}(t)$ for each service client,
which represents the amount of tasks offloaded but not been executed in all edge servers.
In the \emph{t}th time slot, let us denote the amount of tasks of the \emph{i}th service client that has been offloaded but not executed by the \emph{j}th edge server as $H_{i,j}(t)$,
where $H_{i}(t) = \sum_{j\in G_{i}}H_{i,j}(t)$,
the amount of tasks of the \emph{i}th service client which has been
executed by the edge server as $D_{s,i}(t)$.
\begin{equation}
  \centering
  H_{i}(t+1) = max\{H_{i}(t) - D_{s,i}(t), 0\} + D_{r,i}(t), i \in U.
  \label{equ2}
\end{equation}

\subsection{Local Execution and Task Offloading Model}

\indent The \emph{i}th service client's device executes 1bit task
using $L_{i}$ (cycles/bit) CPU cycles \cite{miettinen2010energy, han2016frontal}.
The dynamic frequency and voltage scaling (DVFS) technique is used to adjust
the CPU-cycle frequency of service clients' devices \cite{mao2017survey,mao2017stochastic,tran2017joint}.
In the \emph{t}th time slot, $f_{l,i}(t)$ is the CPU-cycle frequency of the \emph{i}th service client's device.
\begin{equation}
  \centering
  D_{l,i}(t) = \tau f_{l,i}(t) L_{i}^{-1}, i \in U.
  \label{equ3}
\end{equation}
\indent According to circuit theories \cite{chen2015decentralized, wen2012energy},
the CPU power consumption of the \emph{i}th service client's device in
the \emph{t}th time slot is calculated as
\begin{equation}
  \centering
  p_{l,i}(t) = k_{mod,i} f_{l,i}^{3}(t), i \in U,
  \label{equ4}
\end{equation}
\noindent where $k_{mod,i}$ is the effective switched capacitance of the CPU of the
\emph{i}th service client's device.
$f_{max,i}$ is the maximum CPU-cycle frequency of the \emph{i}th service client's device, i.e.,
\begin{equation}
  \centering
  0 \leq f_{l,i}(t) \leq f_{max,i}, i \in U.
  \label{equ5}
\end{equation}
\indent  In this paper, the operational frequency band of the \emph{j}th
edge server is divided equally into $|Z_{j}|$ which $B_{j} = \omega / |Z_{j}|$. In the \emph{t}th
time slot, let us denote the fading of wireless channels
between the \emph{i}th service client's device and the \emph{j}th edge server
as $\gamma_{i,j}(t)$, the channel power gain
from the \emph{i}th service client's device to the \emph{j}th edge server as
\begin{equation}
  \centering
  \Gamma_{i,j}(t) = \gamma_{i,j}(t) g_{0} (\frac{d_{0}}{d_{i,j}})^{\theta}, i \in U, j \in S,
  \label{equ6}
\end{equation}
\noindent where $g_{0}$ is a constant, $\theta$ is an exponent, $d_{0}$ is
the reference distance \cite{mao2017stochastic}, and $d_{i,j}$ is the
distance between the \emph{i}th service client's device and the \emph{j}th edge
server. The task offloading variables in the \emph{t}th time
slot are defined as $\textbf{X}(t)= \{x_{i,j}(t) | i \in U, j \in S\}$,
where $x_{i,j}(t) \in \{0,1\}$, where $x_{i,j}(t) = 1$ indicates that the \emph{i}th
service client's device offloads its computation tasks to the \emph{j}th edge server
in the \emph{t}th time slot.
\begin{equation}
  \centering
  \sum_{j \in S} x_{i,j}(t) \leq 1, i\in U.
  \label{equ7}
\end{equation}
\indent The transmit power of the \emph{i}th service client's device in the \emph{t}th
time slot is denoted by $p_{r,i}(t)$. According to the Shannon-Hartley
formula \cite{cover2012elements}, the transmit rate between the \emph{i}th
service client's device and the \emph{j}th edge server in the \emph{t}th time slot
is
\begin{equation}
  \centering
  r_{i,j}(t) =
  \left\{
    \begin{matrix}
      & B_{j} \log_{2}(1+\frac{\Gamma_{i,j}(t) p_{r,i}(t) }{B_{j}N_{0}}), x_{i,j}=1  & \\
      & 0, x_{i,j} = 0 &
    \end{matrix}
  \right.
  \label{equ8}
\end{equation}
\indent Without loss of generality, $p_{r,i}(t)$ must not exceed the maximum transmit power of the \emph{i}th service client's device.
\begin{equation}
  \centering
  0 \leq p_{r,i}(t) \leq p_{max,i}, i \in U.
  \label{equ9}
\end{equation}
\indent Based on the above definitions, the number of offloading tasks of
the \emph{i}th service client's device is calculated as
\begin{equation}
  \centering
  D_{r,i}(t) = \sum_{j \in S} r_{i,j}(t) \tau , i \in U.
  \label{equ10}
\end{equation}
\subsection{Edge Server Scheduling}
\indent As shown in Fig. \ref{VirtualQueue}, the edge server can transfer input
data to a neighboring edge server via an X2 link \cite{ndikumana2018joint}. The amount of tasks of the \emph{i}th service client's device that have been
executed by the \emph{j}th edge server in the \emph{t}th time slot is denoted as
$D_{s,i,j}(t)$. Then, $D_{s,i}(t)$ can be expressed as
\begin{equation}
  \centering
  D_{s,i}(t) = \sum_{j \in S} D_{s,i,j}(t) , i \in U.
  \label{equ11}
\end{equation}
\noindent For the \emph{j}th edge server, $f_{max,j}$ is the maximum CPU-cycle frequency and $\varphi_{j}$ is the number of CPUs.
\begin{equation}
  \centering
  \sum_{j \in S} D_{s,i,j}(t) L_{i} \leq \tau \varphi_{j} f_{max,j}.
  \label{equ12}
\end{equation}

\section{Problem Formulation}

\indent This section first defines the service cost function and then formulates the average service capacity, i.e., the long-term average number of service clients that can offload their tasks. Finally, the optimization problem studied in this paper is formulated.

\indent The service cost function is defined as follows:
\begin{equation}
  \centering
  \xi_{i}(t) = \beta \xi_{i}^{latency}(t) + (1-\beta) \xi_{i}^{power}(t), i \in U,
  \label{equ13}
\end{equation}
\noindent where $\beta \in [0,1]$. $\xi_{i}^{latency}(t)$ and $\xi_{i}^{power}(t)$
are the latency cost and power cost of the \emph{i}th service client's device in the \emph{t}th time slot.
According to Little’s Law, the average execution delay
of a service client's device is proportional to the average amount of its tasks in the edge environment.
\begin{equation}
  \centering
  \xi_{i}^{latency}(t) = \alpha (Q_{i}(t) - D_{\sum,i}) + (1-\alpha)(H_{i}(t)-D_{s,i}(t)).
  \label{equ14}
\end{equation}
\noindent where $\alpha \in [0,1]$. Therefore, $\xi_{i}^{power}(t)$ is
calculated as
\begin{equation}
  \centering
  \xi_{i}^{power}(t) = p_{l,i}(t) + p_{r,i}(t), i \in U.
  \label{equ15}
\end{equation}

Let us denote the long-term average number of service clients that can offload their tasks as
$\bar{O}$.
\begin{equation}
  \centering
  \bar{O} = \lim_{T \to +\infty }\frac{1}{T \cdot m}\sum_{t=0}^{T-1}\sum_{i \in U, j \in S} x_{i,j}(t).
  \label{equ16}
\end{equation}
\indent Constraint (\ref{equ17}) requires that the service clients' task
queues are stable \cite{neely2010stochastic}.
\begin{equation}
  \centering
  \lim_{T \to +\infty} \frac{\textbf{E}[|Q_{i}(T)|]}{T} = 0,
  \lim_{T \to +\infty} \frac{\textbf{E}[|H_{i}(T)|]}{T} = 0.
  \label{equ17}
\end{equation}

\indent Let us denote the system operation at the \emph{t}th time slot with
$\textbf{R}(t) \triangleq \{\textbf{f}(t),\textbf{p}(t),\textbf{X}(t),\textbf{D}(t)\}$, in
which
$\textbf{f}(t) \triangleq  \{f_{l,i}(t)| i \in U \}$,
$\textbf{p}(t) \triangleq  \{p_{r,i}(t)| i \in U \}$,
$\textbf{D}(t) \triangleq  \{D_{s,j}(t)| j \in S \}$.
Accordingly, the objective functions that minimize the average service cost and maximize the average service capacity respectively are defined below as $\emph{\textbf{P}}_{1}$ together:
\[
  \begin{split}
     \emph{\textbf{P}}_{1}: & \min_{\{\emph{\textbf{R}}(t)\}}  \lim_{T \to +\infty } \frac{1}{T} \sum_{T-1}^{0} \textbf{E}\left [  \sum_{i \in U} \xi_{i}(t) \right ] \\
    & \max_{\{\emph{\textbf{R}}(t)\}}  \bar{O} \\
    & s.t.  (\ref{equ5})(\ref{equ7})(\ref{equ9})(\ref{equ12})(\ref{equ17}), t \in T
  \end{split}
\]

\section{Algorithm Design}

This section introduces an online algorithm for solving the joint task
offloading and resource allocation problem $\emph{\textbf{P}}_{1}$. Based on Lyapunov
optimization, the stochastic optimization problem is converted into a deterministic
optimization problem.

\subsection{Lyapunov Optimization-based Online Algorithm}

We employ the Lyapunov optimization to ensure the stability of the task queues through minimizing the average service cost. To model the problem as a Lyapunov optimization problem, we define the Lyapunov function as:
\begin{equation}
  \centering
  L( \bm{\Theta}(t)) = \frac{1}{2} \sum_{i \in U}[Q_{i}^{2}(t) + H_{i}^{2}(t)].
  \label{equ18}
\end{equation}
\noindent where $\bm{\Theta}(t) = [\textbf{Q}(t),\textbf{H}(t)]$. Then, the
conditional Lyapunov drift is defined as
\begin{equation}
  \centering
  \Delta(\bm{\Theta}(t)) = \textbf{E}[(L(\bm{\Theta}(t+1))-L(\bm{\Theta}(t)))|\bm{\Theta}(t)].
  \label{equ19}
\end{equation}

The Lyapunov drift-plus-penalty function is defined as:

\begin{equation}
  \centering
  \Delta_{\nu}(\bm{\Theta}(t)) =  \Delta(\bm{\Theta}(t)) + V \cdot \textbf{E}[\bm{\xi}(t)|\bm{\Theta}(t)],
  \label{equ20}
\end{equation}
\noindent where $ \bm{\xi}(t) = \sum_{i \in U} \xi_{i}(t) $ and
$ V \in (0,+\infty) $ is a control parameter to keep the
balance between task queues and service cost.

The upper bound of $\Delta(\bm{\Theta}(t))$ with constraints (\ref{equ5}), (\ref{equ7}),\\ (\ref{equ9}) and (\ref{equ12}) is defined as:
\begin{equation}
  \centering
  \begin{split}
    \Delta_{\nu}(t) \leq & C - \textbf{E}[\sum_{i \in U}Q_{i}(t)(D_{\sum,i}(t)-A_{i}(t))|\bm{\Theta}(t)] \\
    & - \textbf{E}[\sum_{i \in U}H_{i}(t)(D_{s,i}(t)-D_{r,i}(t))|\bm{\Theta}(t)] \\
    & + V \cdot \textbf{E}[\bm{\xi}(t)|\bm{\Theta}(t)],
  \end{split}
  \label{equ21}
\end{equation}
\noindent where $\emph{C}$ is a constant \cite{mao2017stochastic}.

According to Lyapunov optimization, in order to minimize $\bm{\xi}(t)$ and maintain the stability of service clients' task queues, we need to minimize the upper bound of $\Delta_{\nu}(t)$ in each time slot as expressed in formula (\ref{equ21}). This minimization is denoted as $\emph{\textbf{P}}_{\textbf{PTS}}$. All the constraints of $\emph{\textbf{P}}_{1}$ except constraint (\ref{equ17}) are included in $\emph{\textbf{P}}_{\textbf{PTS}}$. Algorithm \ref{ag1} presents the pseudo code of the Online Joint Task Offloading and Resource Allocation Algorithm (OJTORA). The pseudo-code for solving $\emph{\textbf{SP}}_{1}$, $\emph{\textbf{SP}}_{2}$ and $\emph{\textbf{SP}}_{3}$ are presented below in Section \ref{sec:optimal_solution}.
\begin{algorithm}[!ht]
  \caption{Online Joint Task Offloading and Resource Allocation (OJTORA)}
  \label{ag1}
  \begin{algorithmic}[1]
    \Require At the beginning of the \emph{t}th time slot, the $\{\Gamma_{i,j}(t)\}$  and $\{A_{i}(t)\}$ will be given.
    \State Get optimal $\textbf{f}(t),\textbf{p}(t),\textbf{X}(t) and \textbf{D}(t)$ by solving
    \[\begin{split}
      \emph{\textbf{P}}_{PTS}: & \min_{\textbf{R}(t)}
      - \sum_{i \in U} Q_{i}(t) D_{\sum,i}(t) \\
      & - \sum_{i \in U} H_{i}(t) (D_{s,i}(t) - D_{r,i}(t))
      + V \cdotp \bm{\xi}(t) \\
      & s.t. (\ref{equ5}) (\ref{equ7}) (\ref{equ9}) (\ref{equ12})
    \end{split}\]
    in which the optimal $\textbf{f}(t),\textbf{p}(t),\textbf{X}(t) and \textbf{D}(t)$
    can be obtained by solving $\emph{\textbf{SP}}_{1}$, $\emph{\textbf{SP}}_{2}$ and $\emph{\textbf{SP}}_{3}$, respectively.
    \State According to the iteration formula in (\ref{equ1}) and (\ref{equ2}), update the $\{Q_{i}(t)\}$ and $\{H_{i}(t)\}$.
    \State Go to the next time slot, which $t=t+1$.
  \end{algorithmic}
\end{algorithm}

\subsection{Optimal Solution to $\textbf{P}_{PTS}$ }
\label{sec:optimal_solution}

\subsubsection{Optimal CPU-Cycle Frequencies of Mobile Devices}

The \emph{$\textbf{SP}_{1}$} is defined as
\[\begin{split}
  \textbf{\emph{SP}}_{1}: & \min_{\{\textbf{\emph{f}}(t)\}}
  \sum_{i \in U}  [-(Q_{i}(t)+V \cdot \alpha \beta)\tau L_{i}^{-1} f_{l,i}(t) \\
  & + V \cdot (1- \beta) k_{mod,i} f_{l,i}^{3}(t)] \\
  & s.t. (\ref{equ5}).
\end{split}\]
\indent The objective function of $\textbf{\emph{SP}}_{1}$ is convex
and constraint (\ref{equ5}) is linear. Therefore, $\textbf{\emph{SP}}_{1}$
is a convex problem. For each service client, we can find that $f_{l,i}(t)$
is independent. Let us denote the optimal CPU-cycle frequency
of the \emph{i}th service client's device as $f_{l,i}^{*}(t)$. We can obtain $f_{l,i}^{*}(t)$ by finding the minimum point of $-(Q_{i}(t)+V \cdot \alpha \beta)\tau L_{i}^{-1} f_{l,i}(t) + V \cdot (1- \beta) k_{mod,i} f_{l,i}^{3}(t)$.
At the same time, $f_{l,i}^{*}(t)$ must fulfil constraint (\ref{equ5}).
\begin{equation}
  \centering
  f_{l,i}^{*}(t) =
  \left\{\begin{matrix}
    & min\{\sqrt{\frac{(Q_{i}(t) + V \cdot \alpha \beta )\tau }{3 k_{mod,i}V(1-\beta )L_{i})}}, f_{max,i}\}, others & \\
    & f_{max,i}, \beta = 1 &
  \end{matrix}\right.
  \label{equ22}
\end{equation}

\subsubsection{Optimal Transmit Power Consumption and Task Offloading Policy}

$p_{r,i}^{*}(t)$ and $\textbf{X}^{*}(t)$ refer to the optimal
$p_{r,i}(t)$ and the optimal  $\textbf{X}(t)$ respectively.
They are obtained by solving \emph{$\textbf{SP}_{2}$}:
\[\begin{split}
  \textbf{\emph{SP}}_{2}: & \min_{\{\textbf{p}(t),\textbf{X}(t)\}} \sum_{i \in U}
   [-\Psi_{i}(t) D_{r,i}(t) + V  (1-\beta) p_{r,i}(t)] \\
  & s.t. (\ref{equ7}) (\ref{equ9}).
\end{split}\]
\noindent When there is $\Psi_{i}(t) = Q_{i}(t) - H_{i}(t) + V \cdot \alpha \beta \leq 0$, the object function of \emph{$\textbf{SP}_{2}$} is non-decreasing with
$p_{r,i}(t)$. Then, because of constraint (\ref{equ9}), $p_{r,i}^{*}(t) = 0$
means that the \emph{i}th service client's device will not offload its tasks to edge servers. Let us denote $U_{off}(t)=\{i|\Psi_{i}(t) \geq 0, i \in U\}$.
The optimal transmit power consumption and the optimal task offloading policy can be achieved by solving \emph{$\textbf{SP}_{2}^{'}$}:
\[\begin{split}
  \textbf{\emph{SP}}_{2}^{'}: & \min_{\{\textbf{p}(t), \textbf{X}(t)\}} \sum_{i \in U_{off}(t)}
   [-\Psi_{i}(t) D_{r,i}(t)
   + V (1-\beta) p_{r,i}(t)] \\
   & s.t. (\ref{equ7}) (\ref{equ9}).
\end{split}\]

This objective function is determined by two variables, i.e., $\textbf{p}(t)$
and $\textbf{X}(t)$. It is hard to obtain $p_{r,i}^{*}(t)$ and $\textbf{X}^{*}(t)$
simultaneously. Therefore, we divide \emph{$\textbf{SP}_{2}^{'}$} into two sub-problems.

\indent First, we assume a $\textbf{X}^{*}(t)$. For a fixed computation
offloading policy, we denote the first sub-problem of \emph{$\textbf{SP}_{2}^{'}$}
as \emph{$\textbf{SP}_{PWR}$}, which can be expressed as
\[\begin{split}
  \textbf{\emph{SP}}_{PWR}: & \min_{\{\textbf{p}(t)\}}  \sum_{i \in U_{off}(t)}
    [-\Psi_{i}(t) D_{r,i}(t)
    + V (1-\beta) p_{r,i}(t)] \\
    & s.t. (\ref{equ9})
\end{split}\]
where $D_{r,i}(t)$ is non-decreasing with $p_{r,i}(t)$. The objective
function of \emph{$\textbf{SP}_{PWR}$} is convex and constraint (\ref{equ9})
is linear. Therefore, \emph{$\textbf{SP}_{PWR}$} is a convex problem.
Similar to \emph{$\textbf{SP}_{1}$}, \emph{$\textbf{SP}_{PWR}$} can be
decomposed for individual mobile devices. Let us denote the edge server
to which the ith service client offloads its computation tasks in the
\emph{t}th time slot as $j^{*}$, there is $x_{i,j^{*}} =1$. In
addition, $r_{i,j^{*}}(t)$ can be obtained by
\begin{equation}
  \centering
    r_{i,j^{*}}(t) = B_{j^{*}} \log_{2} (1 + \frac{\Gamma_{i,j^{*}}(t) p_{r,i}(t)}{B_{j^{*}} N_{0}}),
  \label{equ23}
\end{equation}
where $r_{i,j}(t) = 0$ if $j \neq j^{*}$. Then, $D_{r,i}(t)=r_{i,j^{*}}(t)\tau$.
Therefore, we can obtain $p_{l,i}^{*}(t)$ by finding the minimum of $-\Psi_{i}(t) D_{r,i}(t) + V \cdot (1-\beta) p_{r,i}(t)$. Let us denote $\Lambda_{i}(t) = \frac{\Psi_{i}(t)\tau B_{j^{*}}}{V \cdot (1-\beta)\ln2} - \frac{N_{0} B_{j^{*}}}{\Gamma_{i,j^{*}}(t)}$. Then, the $p_{l,i}^{*}(t)$ can be written as
\begin{equation}
  \centering
  p_{l,i}^{*}(t) =
  \left\{\begin{matrix}
    &min\{\Lambda_{i}(t) , p_{max,i}\}, others  & \\
    & p_{max,i}, \beta = 1 &
  \end{matrix}\right.
  \label{equ24}
\end{equation}

Secondly, we assume a fixed $p^{*}(t)$ and denote the second sub-problem of \emph{$\textbf{SP}_{2}^{'}$} as \emph{$\textbf{SP}_{CO}$}, expressed as follows:
\[
  \begin{split}
    \textbf{\emph{SP}}_{CO}: & \min_{\{\textbf{X}(t)\}}
    \sum_{i \in U_{off}(t)} [-\Psi_{i}(t) D_{r,i}(t)] \\
    & s.t. (\ref{equ7}).
  \end{split}
\]

For a service client where $i \in U_{off}(t)$, it independently
selects an edge server to offload its computation tasks. Therefore, \emph{$\textbf{SP}_{CO}$} can be decomposed for individual service clients. In order to solve \emph{$\textbf{SP}_{CO}$}, we propose an algorithm to obtain $\textbf{X}^{*} (t)$, as described in Algorithm \ref{ag2}.
\begin{algorithm}[!ht]
  \caption{Offloading Server Selection}
  \label{ag2}
  \begin{algorithmic}[1]
    \Require $\varepsilon_{i}(t) = \varnothing, min_{i} = MIN\_DOUBLE$.
    \ForAll{$i \in U$}
      \For{each $j \in G_{i}(t) \& j \notin \varepsilon_{i}(t) $}
        \State $min = \Psi_{i}(t) D_{r,i}(t)$;
        \If{$min_{i} \leq min$}
          \State $j^{*} = j$;
        \EndIf
        \State $G_{i}(t)=G_{i}(t)|\{j\},\varepsilon_{i}(t)=\varepsilon_{i}(t) \cup \{j\} $;
      \EndFor
      \State $x_{i,j^{*}}=1$;
    \EndFor
  \end{algorithmic}
\end{algorithm}

\subsubsection{Computing Resource Allocation of Edge Servers}

All the parts remaining in \emph{$\textbf{P}_{PTS}$} are only
related to the allocation of edge servers' computing resources, defined as \emph{$\textbf{SP}_{3}$}:
\[
  \begin{split}
    \textbf{\emph{SP}}_{3}: & \min_{\{\textbf{D}(t)\}}
  \sum_{i \in U} [-[V(1-\alpha)\beta+H_{i}(t)] \cdot D_{s,i}(t)] \\
  & s.t. (\ref{equ12}).
  \end{split}
\]

The value achieved by the objective function of \emph{$\textbf{SP}_{3}$}
decreases with $D_{s,i}(t)$. Let us denote $value_{i}=V(1-\alpha)\beta+H_{i}(t)$. It can be found that the greater the $value_{i}$,
the more benefit will be generated when an edge server executes
an equivalent amount of tasks. Then, OJTORA employs Algorithm \ref{ag3}
to solve \emph{$\textbf{SP}_{3}$}.
\begin{algorithm}[!ht]
  \caption{Computing Resource Allocation}
  \label{ag3}
  \begin{algorithmic}[1]
    \Require According to the value of $value_{i}$,
    sort mobile devices in a decreasing order.
    $i=1$, $rest_{j}=\tau \varphi_{i} f_{max,j}$.
    \While{$i \leq n$}
      \If{$\sum_{j \in G_{i}(t)}rest_{j} \geq H_{i}(t)$}
        \State $D_{s,i}(t) = H_{i}(t)$;
        \State $rest_{j} = rest_{j} - \frac{rest_{j}}{\sum_{j \in G_{i}(t)}rest_{j}}D_{s,i}(t)$;
      \Else
        \State $D_{s,i}(t) = \sum_{j \in G_{i}(t)}rest_{j}$;
        \State $rest_{j}=0$;
      \EndIf
    \State $i=i+1$;
    \EndWhile
  \end{algorithmic}
\end{algorithm}

\section{Experimental Evaluation}

This section evaluates the performance of OJTORA against two baseline approaches. All the experiments were conducted on a machine equipped with Intel Core i5-7400T processor (4 CPUs, 2.4GHz) and 8GB RAM, running Windows 10 x64.

\subsection{Baseline Approaches}

To our best knowledge, OJTORA is the first attempt to consider both the resource allocation and task offloading in a multi-server edge computing environment for latency-tolerant applications. Due to the issue of edge server coverage overlapping, existing approaches designed for the single-server edge computing environment cannot be directly applied to the multi-server environment. Thus, in the experiments, OJTORA is evaluated against two intuitive baseline approaches, namely Random and Greedy:

\begin{itemize}
  \item \textbf{Random}: The Random approach select edge servers randomly to offload service clients' tasks.
  \item \textbf{Greedy}: In (\ref{equ6}), a shorter distance between an edge
            server and a service client's device results in a smaller communication interference. Based on this fact, the Greedy approach selects the closest edge servers to offload service clients' tasks.
\end{itemize}

\subsection{Experiment Settings}

The experiments are set up in a way similar to \cite{mao2017stochastic}. In each experiment, a total of $\emph{n}$ service clients are distributed in an area covered by $\emph{m}$ base stations. The radius of each base station is 150 m, and there is $\gamma_{i,j}(t)\sim Exp(1), i \in U, j \in S$.
The simulation results are the average values over 10,000 time
slots. In general, there is $\emph{n}$=30, $\emph{m}$=3.
The $A_{i,max} = 1000$ bits.Table \ref{table1}
presents the parameter settings used in the experiments.

\begin{table}[!ht]
  \renewcommand{\arraystretch}{1.3}
  \caption{System Parameters}
  \label{table1}
  \centering
  \begin{tabular}{|c|c||c|c|}
      \hline
      \textbf{Parameter} & \textbf{Value} & \textbf{Parameter} & \textbf{Value}\\
      \hline
      $\omega$  & 10 Hz & $N_{0}$ & -174 dBm/Hz\\
      \hline
      $g_{0}$   & -40 dB & $d_{0}$ & 1 m  \\
      \hline
      $\theta$  & 4      & $k_{mod,i}$ & $1 \times 10^{-27}$ \\
      \hline
      $f_{max,i}$ & 1 GHz & $p_{max,i}$ & 500 mW \\
      \hline
      $L_{i}$   & 737.5 cycles/bit & $f_{max,j}$ & 2.5 GHz\\
      \hline
      $\varphi_{j}$  & 4 & $\tau$ & 2 ms   \\
      \hline
  \end{tabular}
\end{table}

In the experiments, we vary three parameters that may have an impact on OJTORA:

\begin{itemize}
  \item \textbf{Control Parameter $V$:} We change the value of $V$, where $V = 1 \times 10^{9}, 2 \times 10^{9},..., 9 \times 10^{9}$, under four circumstances: 1)
  $\alpha=0.3, \beta=1 \times 10^{-5}$; 2) $\alpha=0.3, \beta=1 \times 10^{-6}$;
  3) $\alpha=0.7, \beta=1 \times 10^{-5}$; and 4) $\alpha=0.7, \beta=1 \times 10^{-6}$.
  \item \textbf{Number of Users $n$:} We change the number of service clients, where $n =10, 20, 30, 100, 200$, with $\alpha=0.3, \beta=1 \times 10^{-5}$,
  $V=1 \times 10^{9}$.
  \item \textbf{Number of Edge Servers $m$:} We change the number of edge servers, where $m = 3, 6, 9$, with $\alpha=0.3, \beta=1 \times 10^{-5}$,
  $V=1 \times 10^{9}$.
\end{itemize}


Two performance metrics are employed to evaluate OJTORA, corresponding to the two optimization objectives.

\begin{itemize}
  \item \textbf{Service Capacity}. Service capacity is measured by the long-term average number of service clients served, as formally defined in (\ref{equ16}).
  \item \textbf{Service Cost}. Service cost is defined as
  $\bar{\xi}_{\sum}=\frac{1}{T \cdot  n} \sum_{t=0}^{T-1} \sum_{i \in U} \xi_{i}(t)$.
  In order to compare the power consumption and service latency more clearly, two
  more specific metrics are defined and employed in the evaluation: 1) the average power consumption of service clients' devices, defined as $\bar{p}_{\sum}=\frac{1}{T \cdot n} \sum_{t=0}^{T-1} \sum_{i \in U} (p_{l,i}(t)+p_{r,i}(t))$; and 2) the average queue length of service clients' devices:
  $\bar{q}_{\sum}=\frac{1}{T \cdot n} \sum_{t=0}^{T-1} \sum_{i \in U} (Q_{i}(t)+H_{i}(t))$.
\end{itemize}

\subsection{Experimental Results}

\subsubsection{Optimality}

\begin{figure}[htbp]
\centering
\subfigure[Average power consumption]{
\includegraphics[width=1.5in]{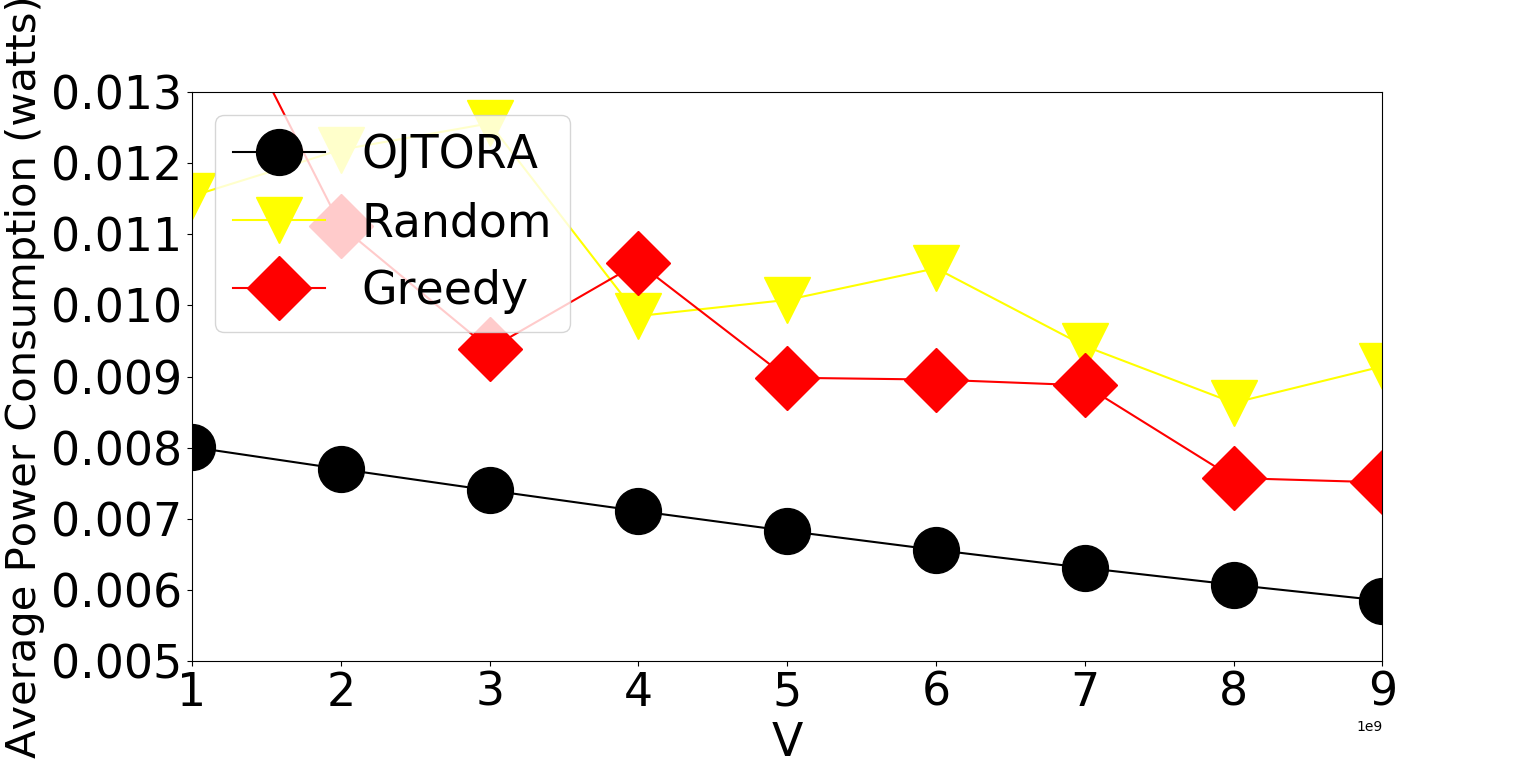}
\label{fig_3_a}
}
\quad
\subfigure[Average queue length]{
\includegraphics[width=1.5in]{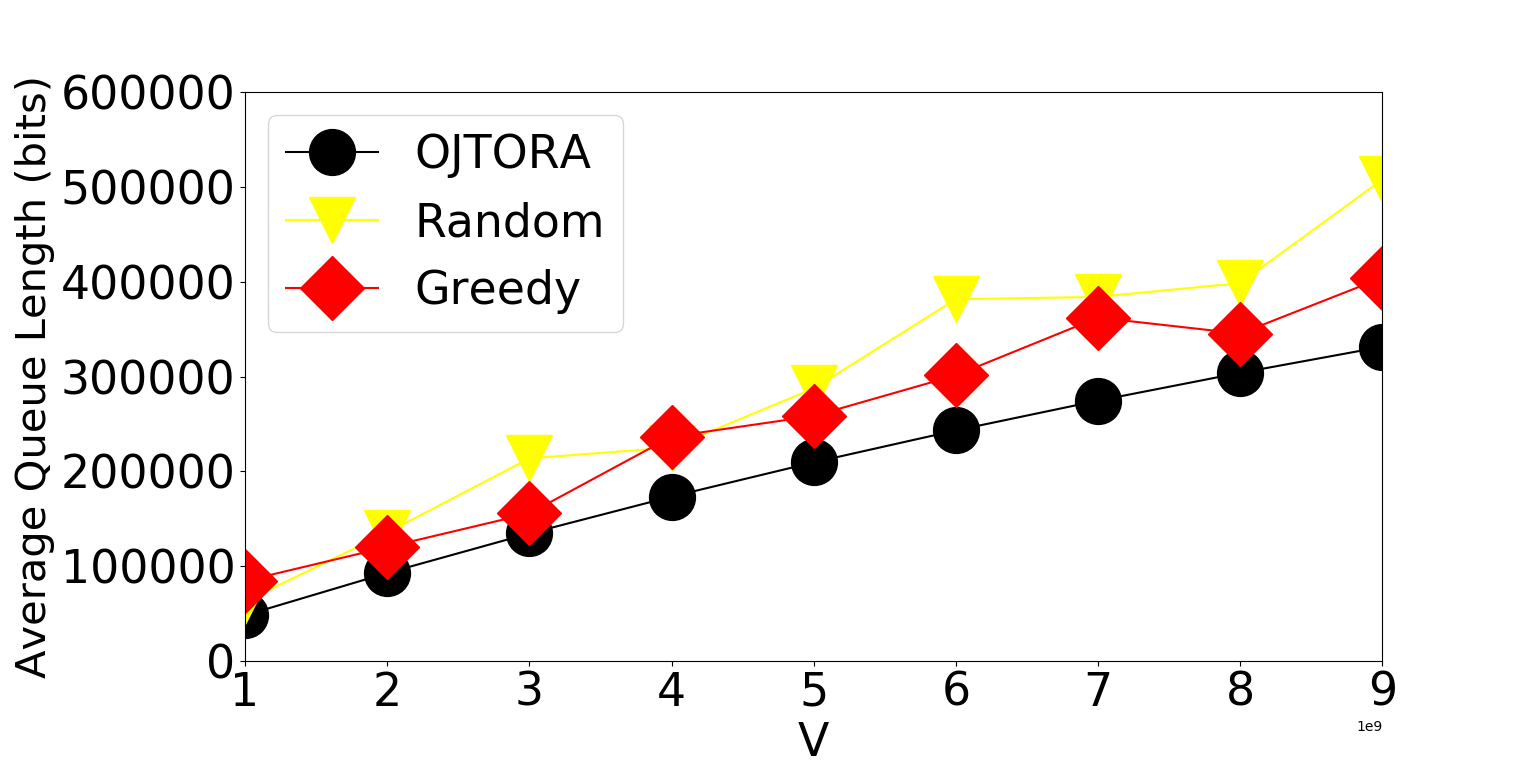}
\label{fig_3_b}
}
\quad
\subfigure[Average service capacity]{
\includegraphics[width=1.5in]{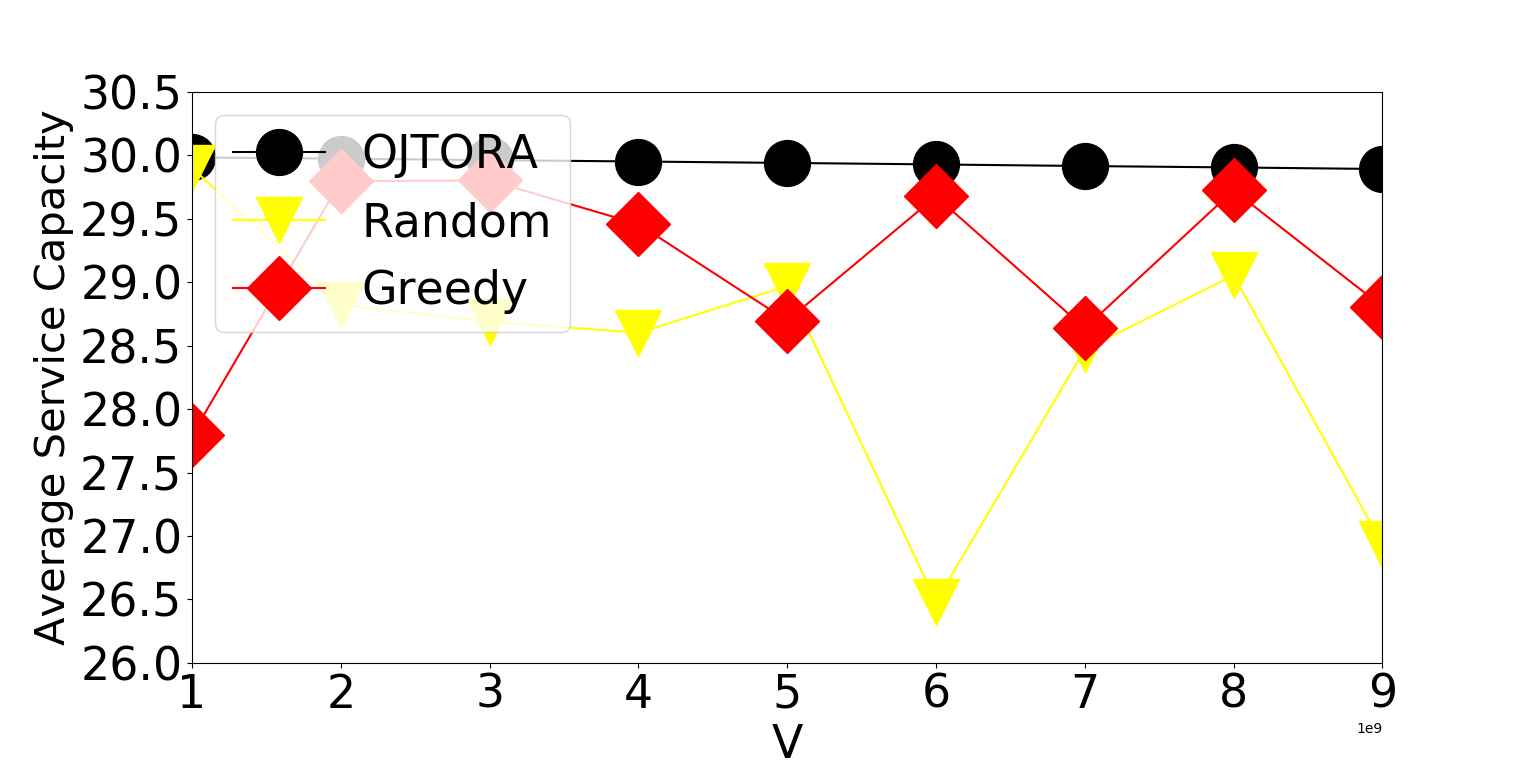}
\label{fig_3_c}
}
\quad
\subfigure[Average service cost]{
\includegraphics[width=1.5in]{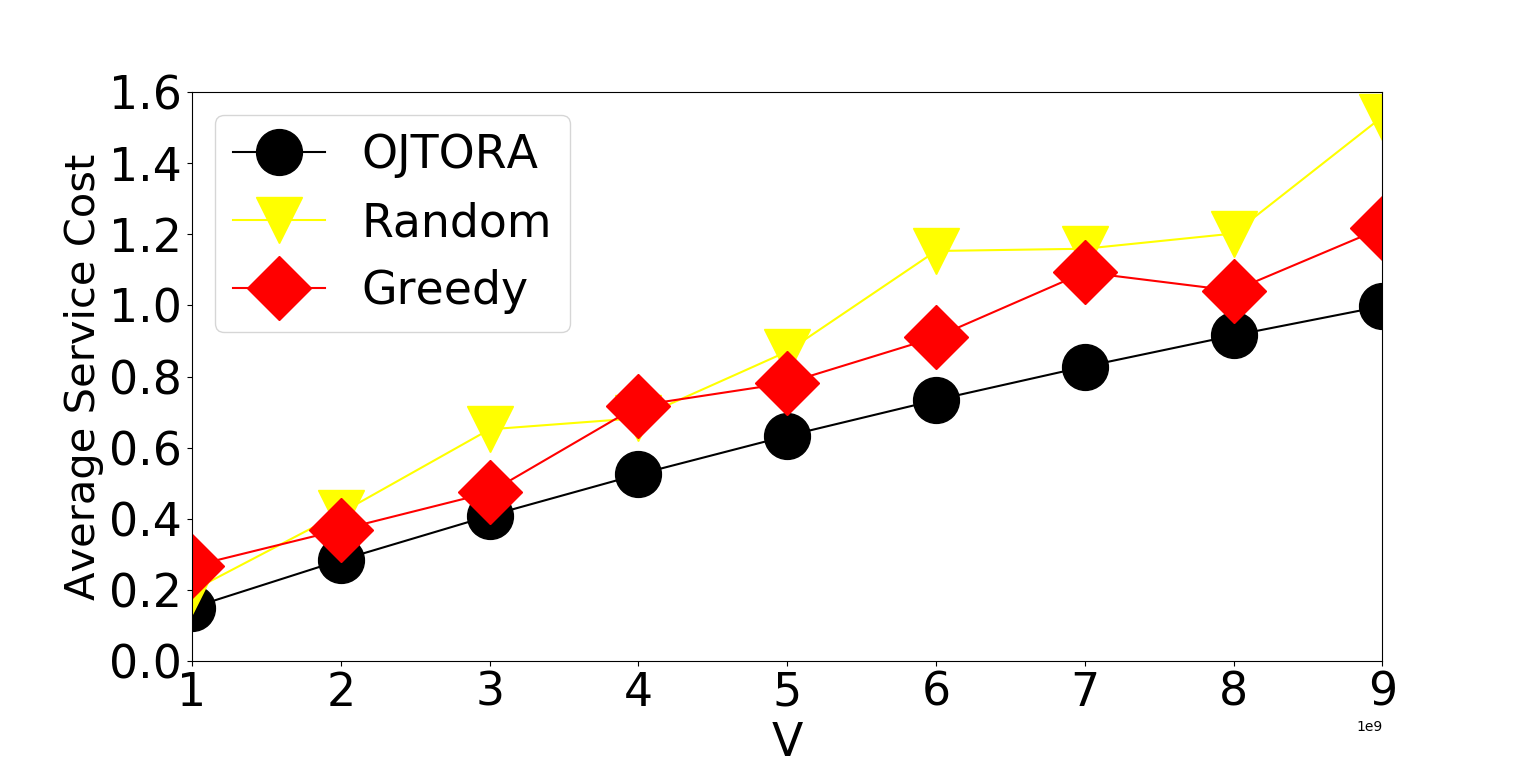}
\label{fig_3_d}
}
\caption{Average power consumption, average queue length and average service cost, average service capacity ($ \alpha=0.3, \beta=1 \times 10^{-5}$)}
\label{fig3}
\end{figure}

As shown in Fig. \ref{fig3}, OJTORA outperforms the two baselines approaches significantly in achieving both optimization objectives. Specifically, it outperforms the Random approach by 27.8\%-41.1\% in power consumption, 23.2\%-37.1\% in queue length, 0.2\%-10.9\% in average service capacity and 23.1\%-37.2\% in average service cost. OJTORA also outperforms the Greedy approach, by 25.6\%-39.1\% in power consumption, 20.1\%-35.7\% in queue length, 0.2\%-6.7\% in average service capacity and 22.1\%-35.8\% in average service cost. Fig. \ref{fig3} also shows that the Greedy approach outperforms the Random approach. The reason is that the Greedy approach selects the closest edge servers to offload service clients' tasks. This reduces the power consumed by the data transmission between service clients' devices and edge servers.


\subsubsection{Impact of Parameter $V$}

\begin{figure}[htbp]
\centering
\subfigure[Average service capacity]{
\includegraphics[width=1.5in]{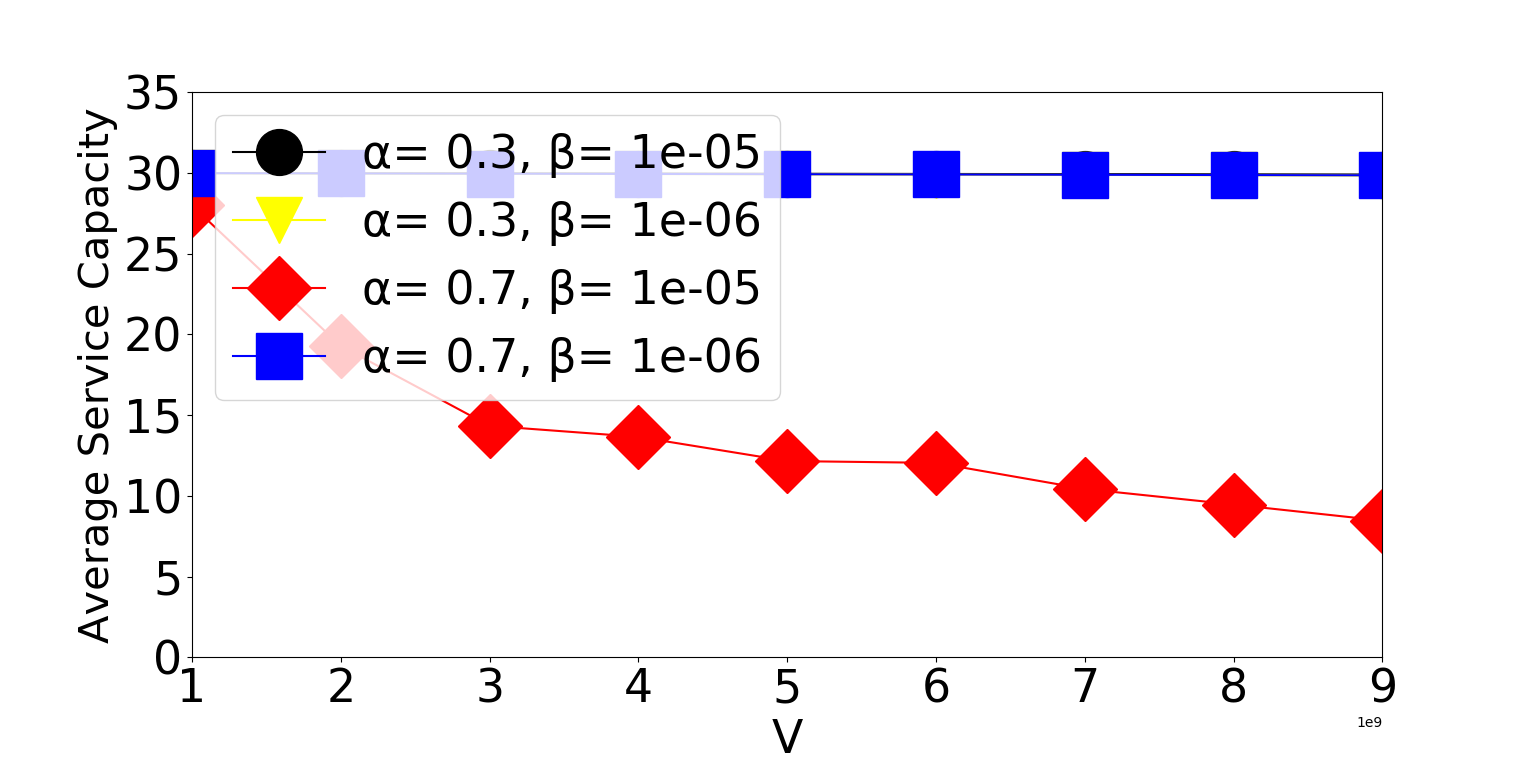}
\label{fig_6_c}
}
\quad
\subfigure[Average service cost]{
\includegraphics[width=1.5in]{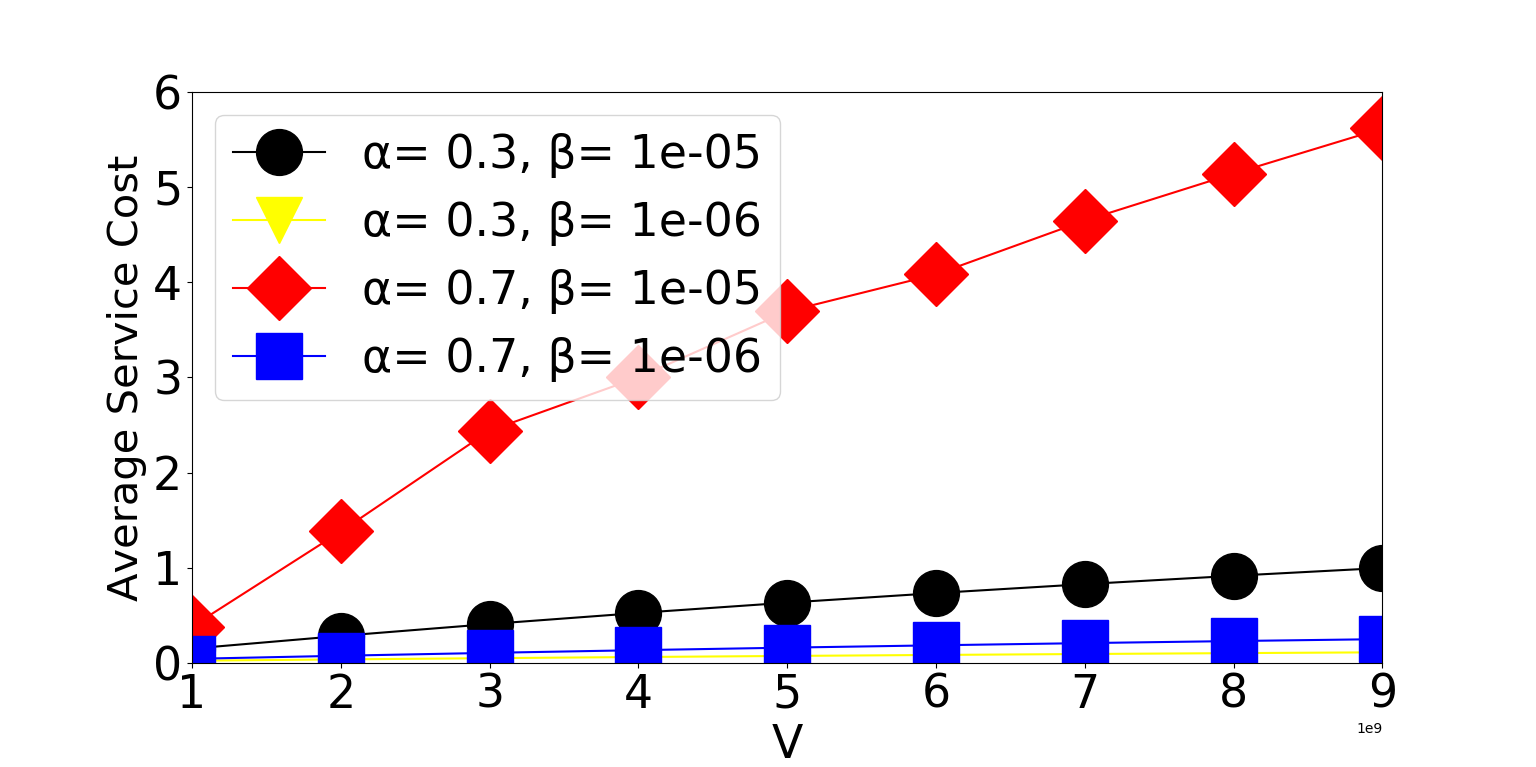}
\label{fig_6_d}
}
\caption{Average service capacity and average service cost vs $V$.}
\label{fig6}
\end{figure}

As shown in Fig. \ref{fig6}, as $V$ increases, the average service capacities achieved by all three approaches decrease. However, the average
service costs increase. According to (\ref{equ22}) and (\ref{equ24}),
as $V$ increases, the transmit power consumption of service clients' devices
decreases. This results in the decrease in the average service capacity. With the increase in V, the power consumption increases more rapidly than the latency. As a result, the average service cost increases.


\subsubsection{Impact of the Number of Service Clients $n$}

\begin{figure}[!ht]
  \centering
  \includegraphics[width=2.5in]{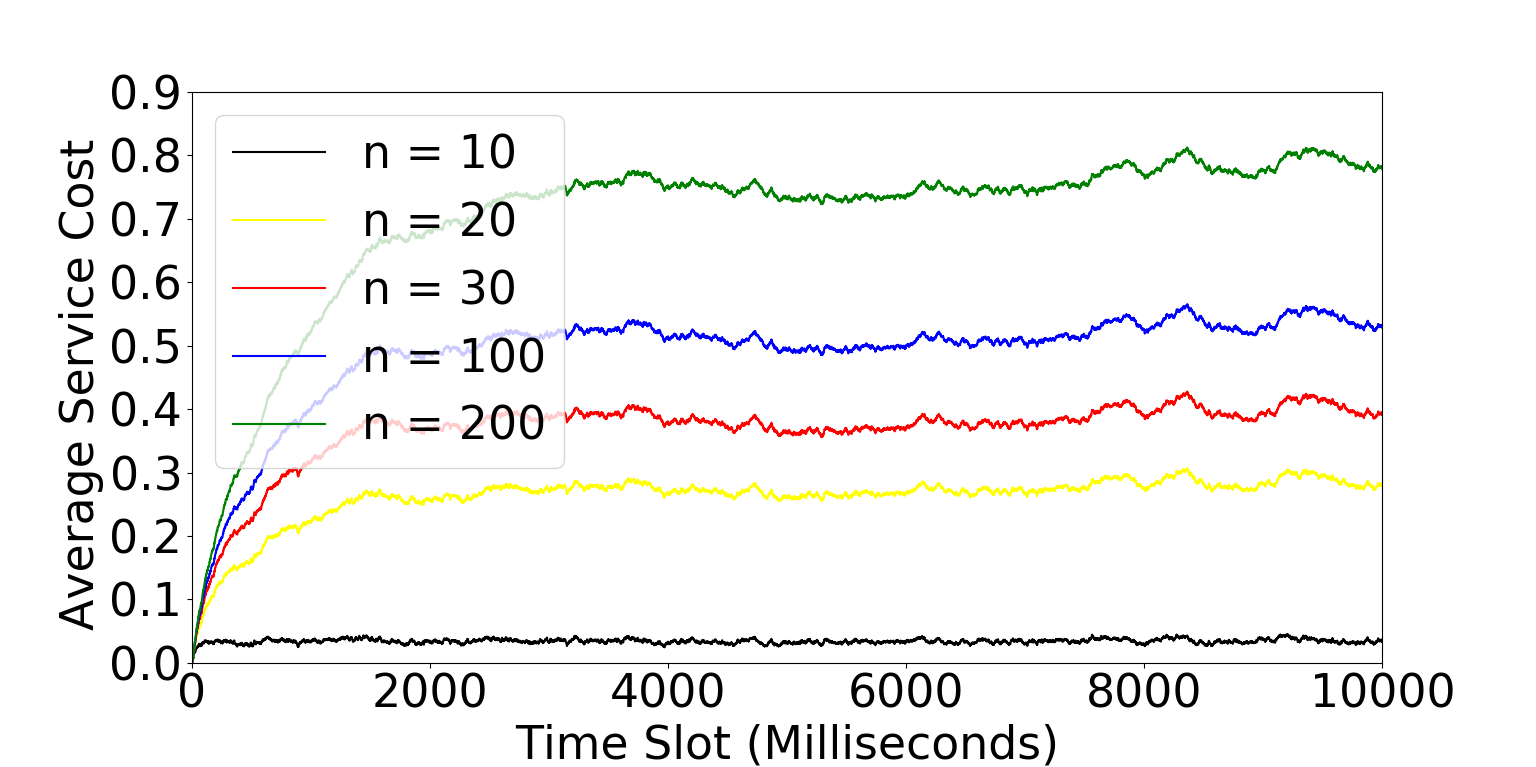}
  \caption{Average service cost per
    time slot vs the number of mobile devices.}
  \label{fig7}
\end{figure}

As shown in Fig.\ref{fig7}, As $n$ increases,
the average service cost increases. This is because the
competition among service clients for the computing resources on the edge servers becomes fiercer with the increase in $n$.

\subsubsection{The Impact of the Number of Edge Servers $m$}

\begin{figure}[!ht]
  \centering
  \includegraphics[width=2.5in]{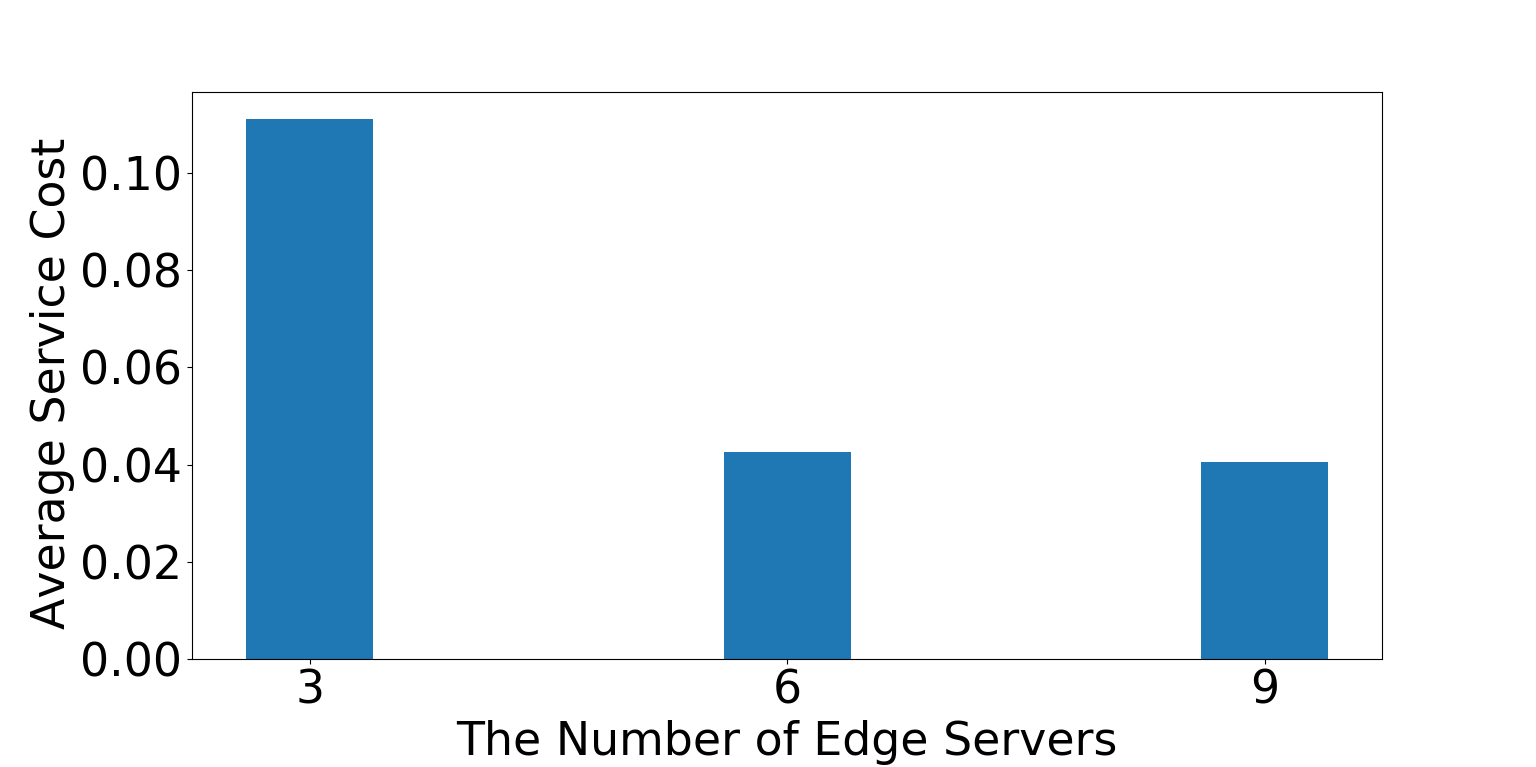}
  \caption{Average service cost vs the number of edge servers.}
  \label{fig8}
\end{figure}

As shown in Fig.\ref{fig8}, with the increase in $m$, the average power consumption, the average queue length and the average service cost increase.
However, when $m$ continues to increase, the average service cost converges and remains steady. This is because the resources are more than sufficient to allow all service clients to be served.

\section{Conclusion}

In this paper, we investigated a joint task offloading and resource allocation problem in multi-server edge computing environments with the objectives to maximize service capacity, i.e., the number of mobile devices served, and to minimize the service cost, i.e., the service latency and power consumption experienced by service clients. To solve this problem, we proposed OJTORA, an online algorithm based on Lyapunov optimization, which converts the stochastic optimization problem to a per-time-slot deterministic optimization problem. The experimental results show that our approach significantly outperforms two baseline approaches.
However, OJTORA does not consider the fairness in the resource sharing among service clients because it assumes static bandwidth allocation for now. In the future, dynamical allocation of bandwidth will be studied. We will also extend OJTORA to accommodate users service clients' mobility.
\bibliographystyle{IEEEtran}
\bibliography{IEEEabrv,IEEEexample}
%
%
%
\end{document}